# Influence of the pseudogap on the Nernst coefficient of $Y_{0.9}Ca_{0.1}Ba_2Cu_3O_y$

M. Matusiak[1,2,*], S.H. Naqib[1,3], I. Kokanović[1,4], J.R. Cooper[1]

[1] *Cavendish Laboratory, Department of Physics, University of Cambridge*

*J.J. Thomson Avenue, Cambridge CB3 0HE, U.K.*

[2] *Institute of Low Temperature and Structure Research, Polish Academy of Sciences,*

*P.O. Box 1410, 50-950 Wrocław, Poland*

[3] *Department of Physics, University of Rajshahi, Raj-6205, Bangladesh*

[4] *Department of Physics, Faculty of Science, University of Zagreb, P.O.Box 331, Zagreb, Croatia*

**PACS**

74.25.Fy, 74.72.Bk, 72.15.Jf

**Abstract**

We have studied the Nernst coefficient, $\nu(T)$, of epitaxial thin films of the superconductor $Y_{1-x}Ca_xBa_2Cu_3O_y$ with $x = 0.05$ and $x = 0.1$. The $x = 0.1$ sample has been measured at three different values of $y$, in the over- (OV), optimally- (OP), and under-doped (UD) states. As the doping level is reduced, $\nu(T)$ starts to fall linearly with $T$ at the temperature ($T^*$) where we expect to see the influence of the pseudogap. The onset temperature of the superconducting fluctuation contribution to $\nu(T)$ was found to vary slowly with the hole concentration ($p$) between $p = 0.118$ and $0.197$. For the OP and UD samples, $\nu(T>T^*)$ is unusually large, being comparable with $|S \tan\theta|$ where $S$ is the Seebeck coefficient and $\theta$ the Hall angle.

**Text**

The origin of the pseudogap in the electronic structure of high-$T_c$ superconductors remains one of the mysteries concerning these compounds. Despite the lack of consensus, one can divide proposed scenarios into two groups: those claiming that the pseudogap is a precursor of the superconducting state [1-3] and those regarding the pseudogap and superconductivity as independent, or even competitive, phenomena [4-7]. Measurements of the Nernst effect could be a good method for investigating the pseudogap state. It is sometimes referred to as the thermal analogue of the Hall effect Both coefficients are measured with similar sample geometry, but for the Nernst effect the voltage signal is a result of the flow of the thermal, not electrical, current. In the normal metallic state the size of $\nu$ can often be very small due to the Sondheimer cancellation [8,9]. In case of a single parabolic band and a momentum ($\underline{k}$) and energy ($\varepsilon$) independent relaxation time ($\tau$) the cancellation is exact and $\nu$ equals to zero. On the other hand, in the mixed state of a superconductor, movement of the vortices produce a significant voltage [10], therefore the Nernst coefficient below $T_c$ can be a few orders of magnitude larger than in the normal state [11-13]. Thus, even the residual presence of vortex-like excitations in the normal state should be easily detectable experimentally. In fact, $T_{onset}$, the temperature where $\nu$ starts to deviate from a small, "normal" value, is much higher than $T_c$ for the high-$T_c$ superconductor $La_{2-x}Sr_xCuO_4$ [9,14], and the $T_{onset}(x)$ dependence resembles the line that is usually drawn for $T^*(x)$ in the "precursor" scenario. These results were interpreted as a sign that the anomalous Nernst effect and pseudogap are controlled by the same energy scale. On the other hand, it has to be emphasized that other work suggests no such correlation [15]. Recently, on the basis of more detailed measurements of $\nu(T)$ for similar Y(Ca)BCO and Zn substituted Y(Ca)BCO films nearer $T_c$, we have argued that most of our data and those for UD $La_{2-x}Sr_xCuO_4$ [10,14] are dominated by weak (Gaussian) superconducting fluctuations and not by the pseudogap [16].

In the present paper we mainly focus on the Nernst effect well above $T_c$. We do not see any relation between a temperature region where $\nu$ is anomalously enhanced, and the value of $T^*$ - the crossover temperature below which signs of a pseudogap appear. On the contrary there is a clear tendency for $\nu(T)$ to decrease below $T^*$. One intriguing possibility is that there is greater electron-hole symmetry below $T^*$ and the results are connected with observations of electron-hole pockets in recent high-field studies [17].

High-quality c-axis oriented thin films of $Y_{1-x}Ca_xBa_2Cu_3O_y$ were grown previously [18] on polished (001) $SrTiO_3$ substrates using pulsed laser deposition and high-density, single-phase sintered targets. Samples were characterized [18] by using X-ray diffraction (XRD), atomic force microscopy, ab-plane room-temperature thermopower, and ab-plane resistivity, $\rho_{ab}(T)$, measurements. XRD was used to determine the structural parameters, phase purity, and degree of c-axis orientation (from the rocking-curve analysis). AFM was employed to study the grain size and the thickness of the films. The films used in the present study were phase-pure and had a high-degree of *c*-axis orientation (typically the full width at half-maximum of the (007) peak was 0.20°). The thickness of the films was in the range (2800 ± 300) Å. The deposition temperature and the *in-situ* oxygen partial pressure for the $Y_{0.9}Ca_{0.1}Ba_2Cu_3O_y$ film used in the present study were 800°C and 1.20 mbar, respectively. Further details of their properties can be found in refs. [18,19].

The Nernst effect was measured under isothermal conditions meaning that the transverse temperature gradient is "shorted out" by the substrate of the film. Measurements were made at stabilized temperatures, stepping the magnetic field between -11 T and +11 T. The films were twinned, but because of the presence of Ca they were more O-deficient for a given $p$ and hence contributions from the Cu-O chains are less significant. Figure 1 shows the $T$ dependences of the Nernst coefficient in the normal state for a film with $x = 0.10$ (denoted as "10%"), that had been oxidized to the overdoped state, and a film with $x = 0.05$ (denoted as "5%"), that had been reduced to the underdoped state. For the over-doped film $\nu(T)$ rises monotonically with falling temperature, whereas for the underdoped film $\nu(T)$ starts to fall at $T \sim 180$ K and becomes negative at $T \sim 120$ K. The temperature, where this fall begins, correlates well with $T^*$ from the resistivity ($\rho$) shown in the inset in Fig. 1. The characteristic downturn in the $\rho(T)$, visible in the "5%" film at $T^* \sim 200$ K is usually recognized as a manifestation of pseudogap [20,21]. We do not observe any similar behaviour in the $\rho(T)$ dependence of the 10% film. It stays linear down to $T \sim 110$ K, below which the rapid and accelerating downturn is ascribed to superconducting fluctuations, in agreement with the analysis in Ref. [22]. To verify whether the behavior of $\nu(T)$ below $T^* \sim 200$ K for x=0.05 is related to the pseudogap, we decided to reduce the oxygen content of the $x$=0.1 film gradually. It was initially overdoped (OV, $T_c$ = 82.2 K), then we obtained the nearly optimal (OP, $T_c$ = 84.5 K), followed by an underdoped (UD, $T_c$ = 81.7 K) composition. Figure 2 shows the temperature dependences of the resistivity for the $x$ = 0.1 film at the three levels of oxidation, where arrows locate the temperature where $\rho(T)$ curves start to fall below the high-$T$ line. In figure 3 we show $\nu(T)$ for the three different dopings, as well as estimates of $S \tan\theta$, where $S$ is the Seebeck coefficient and $\theta$ the Hall angle. As can be seen in the top panel of figure 3, the normal state $\nu(T)$ dependence for the OV sample does not show any specific feature. At lower $T$, $\nu$ grows monotonically and below $T \sim 110$ K the growth becomes more rapid, which seems to be related to the presence of superconducting fluctuations that are also evident in $\rho(T)$. It is worth mentioning that this is the first time that the Nernst coefficient has been measured for a significantly overdoped Y123 sample. This is important because several experimental studies have found indications that pseudogap is absent in high-$T_c$ cuprates with $p > 0.19$ [4-6].

We do not observe a similar growth of the Nernst coefficient in OP and UD samples when $T$ falls below $T^*$, which according to the $\rho(T)$ data in Fig. 2 are $\sim 150$ K (OP) and $\sim 200$ K (UD). Instead the value of $\nu$ starts to drop near $T^*$ and then rises rapidly below $T \sim 110$ K. In the presence of superconducting fluctuations one could expect to see an enhancement of the Nernst coefficient, but surprisingly the pseudogap seems to suppress $\nu$.

The normal state Nernst coefficient is composed of two terms [9]:

$$\nu = \frac{E_y}{(-\nabla_x T)B} = \left[\frac{\alpha_{xy}}{\sigma} - S\tan\theta\right]\frac{1}{B}, \qquad (1)$$

(where $\alpha_{xy}$ is the off-diagonal Peltier conductivity $J_y = \alpha_{xy}(-\nabla_x T)$, and the Hall angle $\theta = \sigma_{xy}/\sigma$).

In this expression the $S\tan\theta$ term arises from the macroscopic condition that there is no transverse electrical current $J_y = 0$ so it is always present for any electronic structure. However for a quadratic energy dispersion $\varepsilon(\underline{k})$ with a momentum ($\underline{k}$) and energy ($\varepsilon$) independent relaxation time $\tau$, the two terms in Eq. 1 cancel exactly [8,9]. When $d\tau/d\varepsilon$ is significant, i.e. $S$ is reasonably large, then the resulting $\nu(T)$ is also of order $S \tan\theta$ [8,23]. However Clayhold [23] has derived the following interesting formula for the case where there is no $\varepsilon$ dependence and $\tau=\tau(\underline{k})$:

$$\nu = [\langle S \tan\theta \rangle - \langle S \rangle \langle \tan\theta \rangle] \frac{1}{B}, \tag{2}$$

Here the $\langle \rangle$ refer to Fermi surface (FS) averages weighted by the electrical conductivity. When the results in Figs. 1 and 3 are scrutinized in the light of the above two formulae the following points emerge. For the OD sample (and only for this sample) the value of $\nu$ is smaller than $|S \tan\theta|$ near 300 K, i.e. more usual metallic behavior is found there. At lower $T$ the value of the $\alpha_{xy}/\sigma$ term in Eq. 1 must $\sim |S \tan\theta|$ since $S$ is negative and $\nu \sim 2|S \tan\theta|$ there. It is possible that these results for the OD film can be understood in terms of Eq. 2 and a rather strong variation of $\tau$ with $\underline{k}$ – the results of such models for most other transport properties have been extensively discussed and reviewed by Hussey [24]. Qualitatively the results for the OD sample suggest that there could be a positive contribution to $S$ from regions of the FS with higher curvature which have larger values of $\theta$.

The results for the OP and UD samples show clearly that even at 300 K – in most cases well above $T^*$, the $\alpha_{xy}/\sigma$ term in Eq. 1 is 2-3 times $S \tan\theta$. So for these two samples $\nu(300)$ is anomalously large and we believe this is caused by the pseudogap since heat capacity and magnetic susceptibility data for $p<0.19$ [25,26] can be interpreted in terms of an energy gap ($E_g$) which does not close or fill in when $T> T^*(=E_g/k_B)$. However for lower $T$, when $T<T^*$, our surprising new result is that $\nu(T)$ falls at a rate comparable with the increase in $S \tan\theta$. In other words the $\alpha_{xy}/\sigma$ term in Eq. 1 seems to become smaller rather quickly. Generally speaking, by analogy with the properties of $\sigma_{xy}$, this could be caused by flat parts of the FS becoming increasingly influential or by increased electron-hole symmetry. However the $\alpha_{xy}/\sigma$ term could also be reduced if large angle scattering processes were frozen out, since it is well known that this would increase the electrical conductivity while having less effect on the thermal conductivity and the Peltier tensor.

Despite not having an exact picture for the normal state, we can extrapolate the T-dependence of the normal state Nernst coefficient ($v_n$) to lower temperatures and check when the additional superconducting component ($v_s$) can be seen. Plots in Fig. 4 show the behaviour with different criteria for $T_{onset}$. If the criterion from Ref. 11, i.e. $v_s = 4$ nV K$^{-1}$ T$^{-1}$, is used, then the $T_{onset}$ is ~ 100 K for all values of $p$ studied and falls slightly as $p$ increases. Similar values of $T_{onset}(p)$ and a similar decrease with $p$ were reported for LSCO and Bi-2212 [11], although for LSCO $T_c$ is lower and therefore $T_{onset} - T_c$ is much larger. However, the precision achieved in our experiment allows us to choose a smaller criterion, e.g. 1 nV K$^{-1}$ T$^{-1}$, or even try to estimate what is temperature of the real onset. Despite inevitable uncertainties in the extrapolation of the normal state component of $v(T)$, last result appears to be surprising, because it shows that $T_{onset}$ rises on the overdoped side of the phase diagram, in contrast to the results for LSCO and Bi-2212. It could mean that inelastic scattering processes, which suppress superconducting fluctuations, are less effective on the overdoped side allowing a longer "tail" in the $v_s(T)$ dependence.

In conclusion, we have studied the temperature dependence of the Nernst coefficient of the $Y_{0.95}Ca_{0.05}Ba_2Cu_3O_y$ ("5%") and $Y_{0.9}Ca_{0.1}Ba_2Cu_3O_y$ ("10%") compounds, where in the second material the oxygen content $y$ was varied to allow us to carry measurements in over- (OV), optimally- (OP) and under-doped (UD) states. Taking into account the known values of $S\tan\theta$ we find that in all samples except the OD one the Nernst coefficient is unusually large at room temperature which we believe could arise from the pseudogap. Then at lower $T$ there is a clear decrease in ν which seems to suggest that the $\alpha_{xy}/\sigma$ term in Eqn. 1 becomes smaller for $T<T^*$. We have not observed a signature of superconducting fluctuations at temperatures higher than ~ 120 K (or in this case $(T-T_c)/T_c = 0.5$) irrespective of the concentration of holes. This could possibly be linked to the effects of inelastic scattering.

**Figure captions**

1. (color online) Temperature dependence of the Nernst coefficient (ν) and $S\tan\theta$ in the $Y_{0.95}Ca_{0.05}Ba_2Cu_3O_y$ ("5%") and $Y_{0.9}Ca_{0.1}Ba_2Cu_3O_y$ ("10%") films. The inset shows data for the thermoelectric power and resistivity of these samples. The dashed lines in the inset show linear extrapolations of the high-T $\rho(T)$. The asterisk in the inset shows the temperature ($T^*$) where $\rho(T)$ for the "5%" sample starts a gradual fall from the linearity.

2. (color online) Temperature dependence of the resistivity of the $Y_{0.9}Ca_{0.1}Ba_2Cu_3O_y$ film in over-doped (OV – red line), optimally doped (OP – green line) and under-doped (UD – blue line) states. The dashed lines show linear extrapolations of the high-T $\rho(T)$. The arrows show the temperatures where $\rho(T)$ starts a gradual fall from the linearity. The inset shows data in the vicinity of superconducting transitions.

3. (color online) Temperature dependence of the Nernst coefficient of the $Y_{0.9}Ca_{0.1}Ba_2Cu_3O_y$ film in the over-doped (OV – the upper panel), optimally doped (OP – the middle panel) and under-doped (UD – the bottom panel) states. The solid lines are guides for eye. The inset in the top panel shows an example of the raw data, measured voltage versus magnetic field, used to calculate ν at a given temperature. The black vertical arrows show the 4 and 1 nV K$^{-1}$ T$^{-1}$ criteria (see text). The dash-dotted curves in the lower two panels show *estimates* of $S\tan\theta$ for the OP and UD samples. These were made using measurements of $S(290)$, knowing that for these $p$ values $S(T)$ curves are parallel to that shown in Fig. 1 for $x = 0.05$, and the fact that $\tan\theta$ is independent of $p$ (e.g. ref. 27).

4. (color online) The variation of the temperature where an anomalous Nernst effect emerges with doping in the $Y_{0.9}Ca_{0.1}Ba_2Cu_3O_y$ film. Three different onset criteria are used 4 nV K$^{-1}$ T$^{-1}$, 1 nV K$^{-1}$ T$^{-1}$, and a "real" one (see text). Uncertainties arising from scatter in the data points and in the normal contribution to $\nu(T)$ are shown by the shaded regions.

**Figure 1**

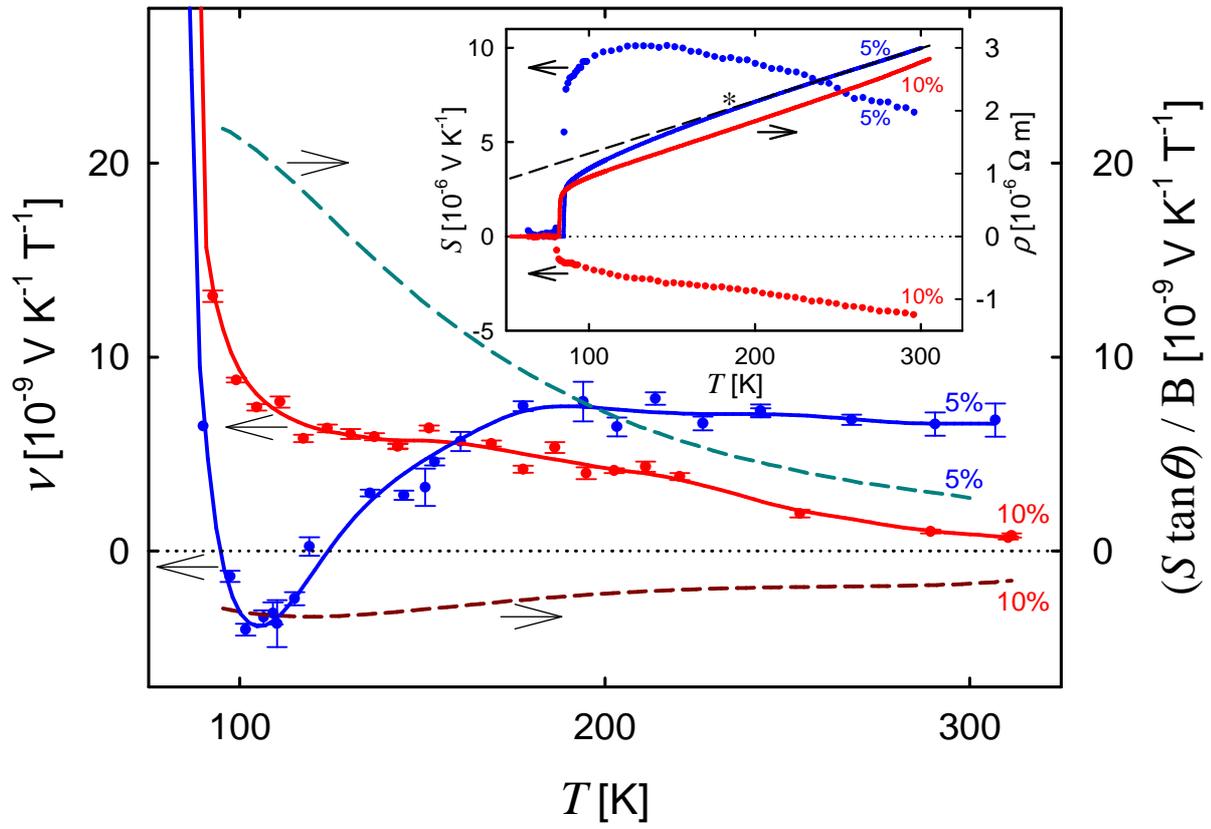

**Figure 2**

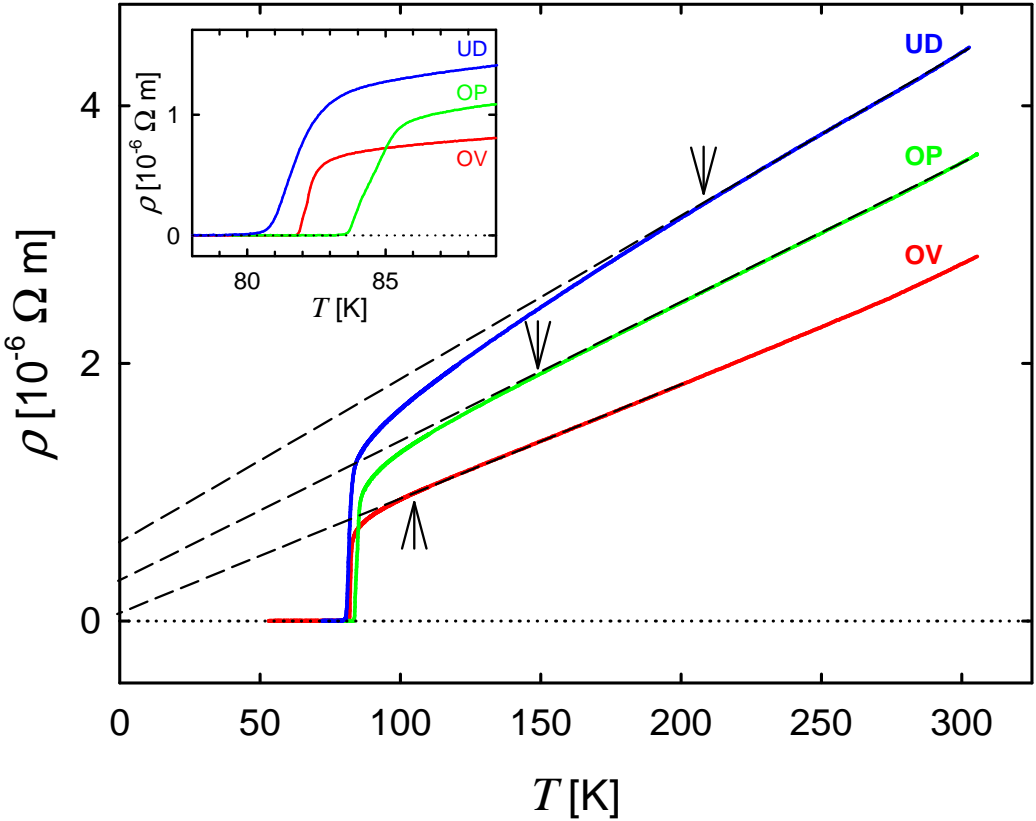

**Figure 3**

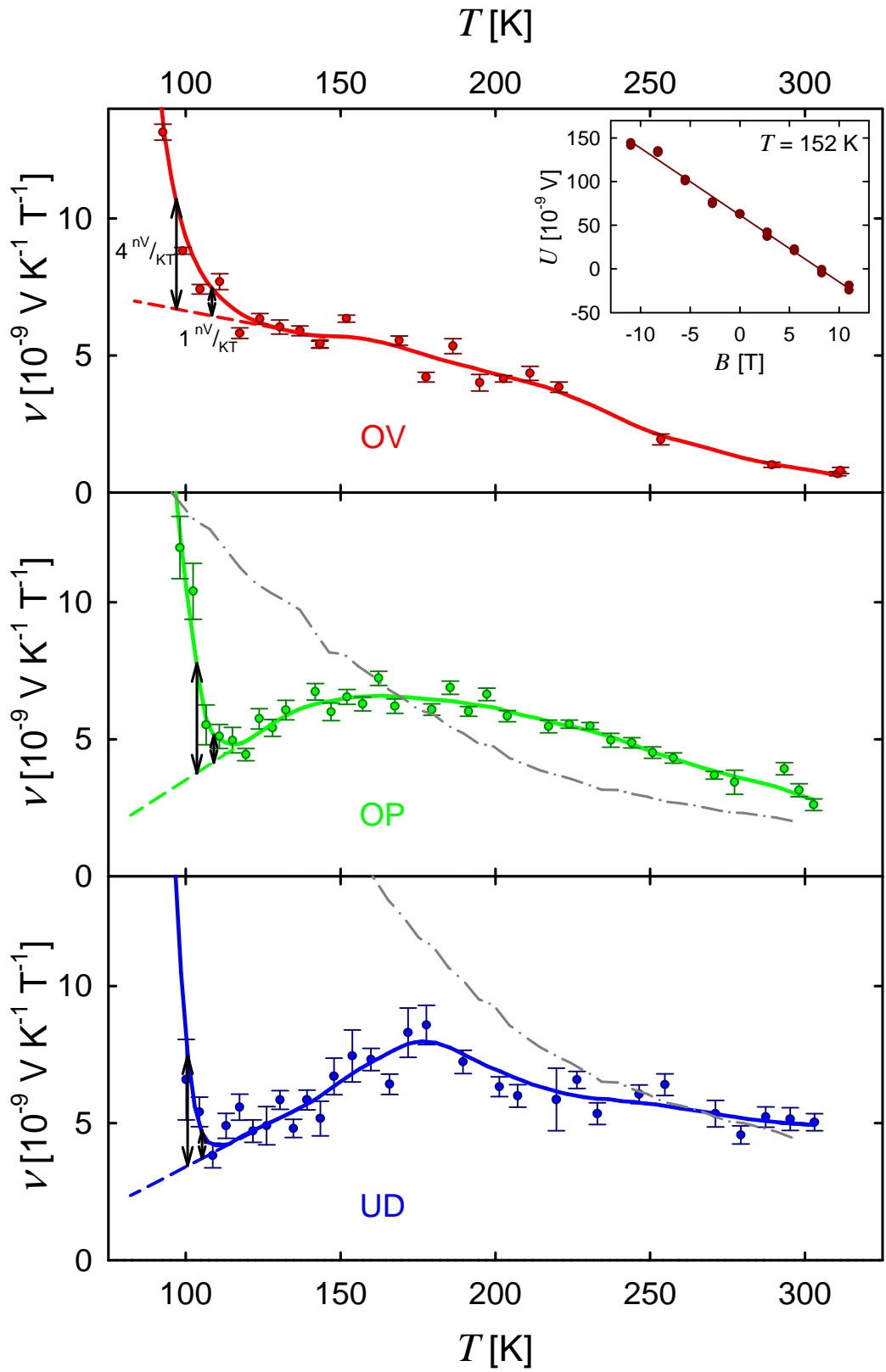

**Figure 4.**

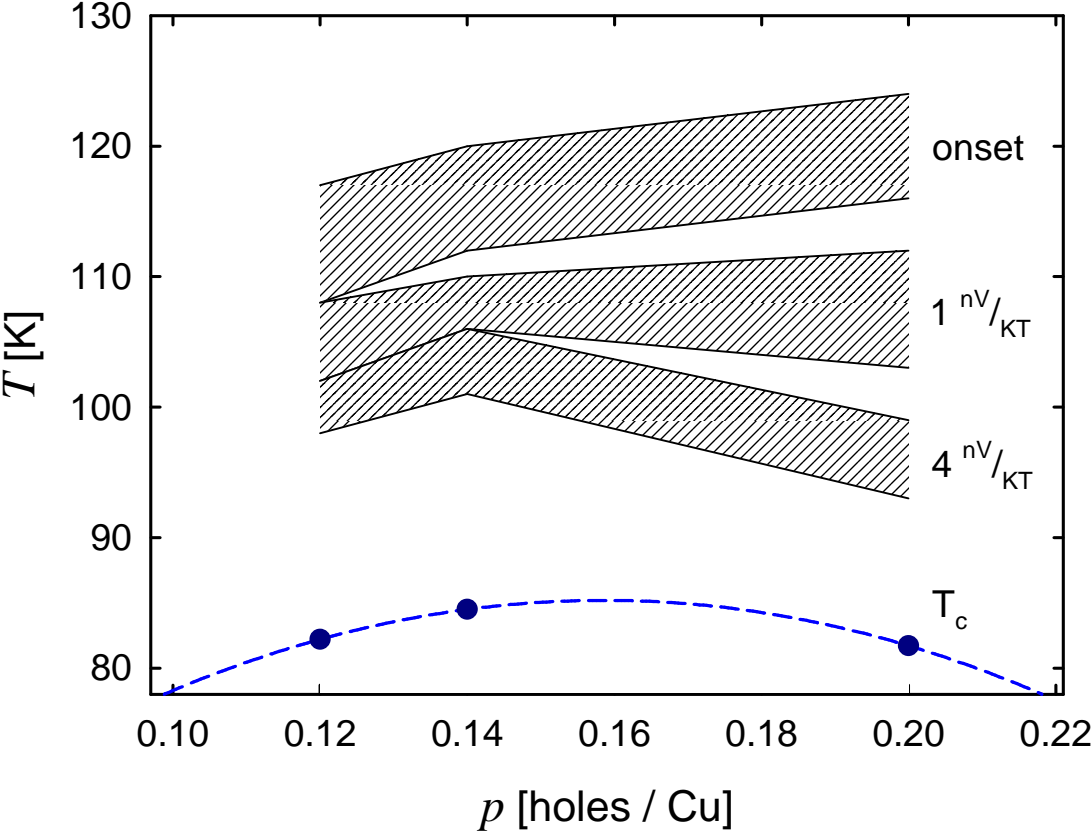